\documentclass[11pt,a4paper]{article}
\usepackage[left=2.2cm,right=2.5cm,top=2.5cm,bottom=2.8cm]{geometry}
\usepackage[utf8]{inputenc}
\usepackage[T1]{fontenc}
\usepackage[table]{xcolor}
\usepackage{amsmath}
\usepackage{amssymb}
\usepackage{amsthm}
\usepackage{mathrsfs}  
\usepackage{graphicx,subcaption}
\usepackage{boldline}
\usepackage{lineno}
\usepackage{tikz}
\usepackage{authblk}
\usepackage{titlesec}
\usepackage{secdot}
\usepackage{hyperref}
\hypersetup{
    colorlinks=true,
    linkcolor=blue,
    citecolor=blue,
    filecolor=magenta,      
    urlcolor=cyan,
}
\usepackage{cleveref}
\titleformat{\section}{\bfseries\large}{\thesection.}{0.5em}{}
\titleformat{\subsection}{\large}{\thesubsection.}{0.5em}{\itshape}
\DeclareRobustCommand{\legendline}[1]{%
  \textcolor{#1}{\rule[0.5ex]{5ex}{0.25ex}}%
}
\DeclareRobustCommand{\legendsquare}[1]{%
  \textcolor{#1}{\rule[0.5ex]{0.75ex}{0.75ex}}%
}

\DeclareMathOperator*{\argmin}{arg\,min}
\title{On the Accuracy of Compressibility Transformations}
\author[1]{M. Engin Danis\thanks{danis@lanl.gov}}
\author[2]{Paul Durbin\thanks{durbin@iastate.edu}}
\affil[1]{{\small Theoretical Division, Los Alamos National Laboratory, Los Alamos, 87545, USA}}
\affil[2]{{\small Department of Aerospace Engineering, Iowa State University, Ames, 50011, USA}}
\begin{document}

\maketitle

\begin{abstract}
This study highlights the importance of satisfying the eddy viscosity equivalence below the logarithmic layer, to deriving accurate compressibility transformations. First, we analyze the ability of known transformations to satisfy the eddy viscosity equivalence and show that the accuracy of these transformation is strongly dependent on this ability. Secondly, in a step-by-step manner, we devise new transformations that satisfy this hypothesis. An approach based on curve fitting of the incompressible Direct Numerical Simulation data for eddy viscosity profiles below the logarithmic layer provides an extremely accurate transformation. That motivates
self-contained methods, making use of mixing length formulas, in the inner region. 

It is shown that the accuracy of existing transformations can be significantly improved by applying these ideas, below the logarithmic layer. Motivated by the effectiveness of the formulations derived from eddy viscosity equivalence, we introduce a new integral transformation based on Reynolds number equivalence between compressible and incompressible flow. This approach is based on defining a new compressible velocity scale, which affects the accuracy of transformations. Several choices for the velocity scale are tested, and in each attempt, it is shown that the eddy viscosity equivalence plays a very important role for the accuracy of compressibility transformations. 

\end{abstract}

\section{Introduction}\label{sec:introduction}
There has been a growing interest in deriving accurate compressibility transformations in the recent years \cite{trettel2016,volpiani2020,Griffin2021,hasan2023,huang2023,hasan2024,cheng2024}. The fundamental objective is to relate compressible flow statistics to those of incompressible flow, in order to facilitate analysis, modeling and prediction of compressible wall-bounded turbulence. Unfortunately, the complete set of conditions required to derive accurate compressibility transformations is unknown. This results in appeals to empiricism, while deriving transformations through the introduction of ad-hoc compressibility terms. Hence, many of the existing methods are unable to guide improvement to predictive models. The present study introduces a new condition for transformations, based on the eddy viscosity scaling, which significantly improves accuracy when applied below the logarithmic layer. 

Compressibility transformations should be derived by treating the constant stress layer and the outer region separately \cite{hasan2024}. In the constant stress layer, Trettel and Larsson argued that profiles of compressible and incompressible shear stresses should collapse under a proper wall-normal coordinate transformation \cite{trettel2016}. In fact, that notion of Reynolds stress equivalence was strengthen by several Direct Numerical Simulations (DNS) of zero-gradient hypersonic flow over a flat plate \cite{duan2010,duan2011,Zhang2018,Huang2020}. Denoting the compressible variables by lowercase letters and incompressible variables by uppercase letters, Trettel and Larsson \cite{trettel2016} reasoned that if the Reynolds stresses follow wall scaling, and the total
stress is constant, viscous stress also scales: 
\begin{equation}\label{eq:vel-grad-hypothesis}
    \frac{\partial U}{\partial Y} = \frac{\mu}{\mu_w}\frac{\partial u}{\partial y}.
\end{equation} 
As pointed out by Danis and Durbin \cite{danis2022}, this results in an eddy viscosity equivalence,
\begin{equation}\label{eq:mut-equivalence}
    \mu_T^+(y) = \mu_{T,i}^+(Y^+), 
\end{equation}
where $\mu_T^+=\mu_T/\mu$ is the nondimensional compressible eddy viscosity as a function of $y$ and $\mu_{T,i}^+=\mu_{T,i}/\mu_w$ is the nondimensional incompressible eddy viscosity as a function of $Y^+$. Here, $\mu=\mu(y)$ is the local dynamic viscosity and $\mu_w$ is its value at the wall $y=0$. This is in fact a very powerful argument that suggests the compressible and incompressible eddy viscosity profiles should collapse under a proper wall-normal coordinate transformation $y\rightarrow Y$. Thus, the function in (\ref{eq:mut-equivalence}) need not be determined from incompressible flow: the general idea of wall-scaling using local variables, $\mu$, $\rho$ and $\tau$ (the local total stress) is  
$$
\mu_T=\mu M(Y^+)
$$
for some function $M$.

As a corollary to \Cref{eq:mut-equivalence}, compressible and incompressible turbulent kinetic energy (TKE) production terms should also be equivalent under the same wall-normal coordinate transformation,
\begin{equation}\label{eq:prod-equivalence}
   \mu P=\mu\overline{\rho u^{''}_iu^{''}_j}S_{ij}= \tau^2\mathcal{P}(Y^+)
\end{equation}
where $\mathcal{P}$ is the wall-scaled production. 

\begin{figure}[htbp]
    \begin{center}
        \begin{subfigure}{0.485\textwidth}
            \centering
            \includegraphics[width=\textwidth,trim={3.4cm 8.5cm 3.4cm 8.5cm},clip]{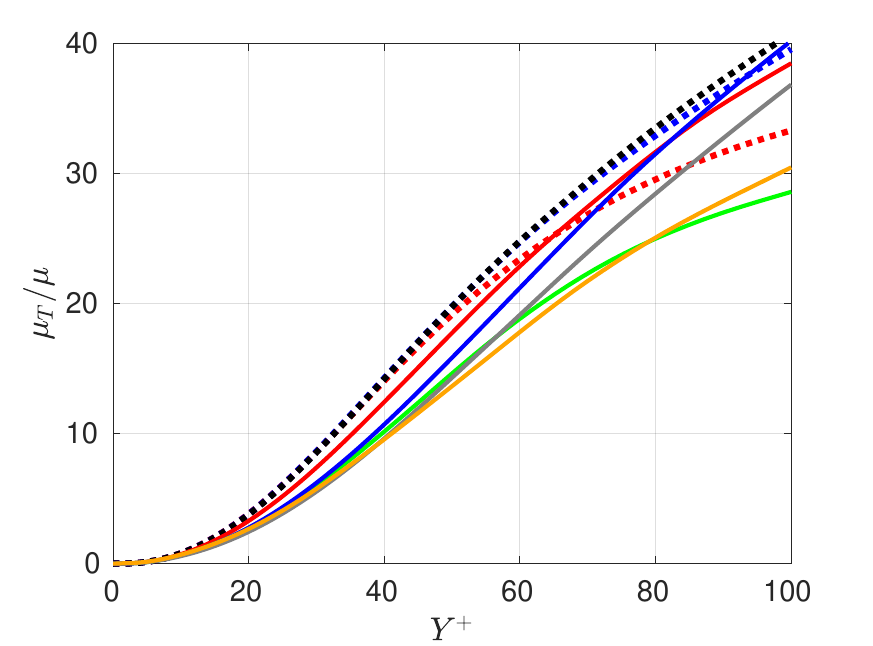}
        \end{subfigure}
        \begin{subfigure}{0.485\textwidth}
            \centering
            \includegraphics[width=\textwidth,trim={3.4cm 8.5cm 3.4cm 8.5cm},clip]{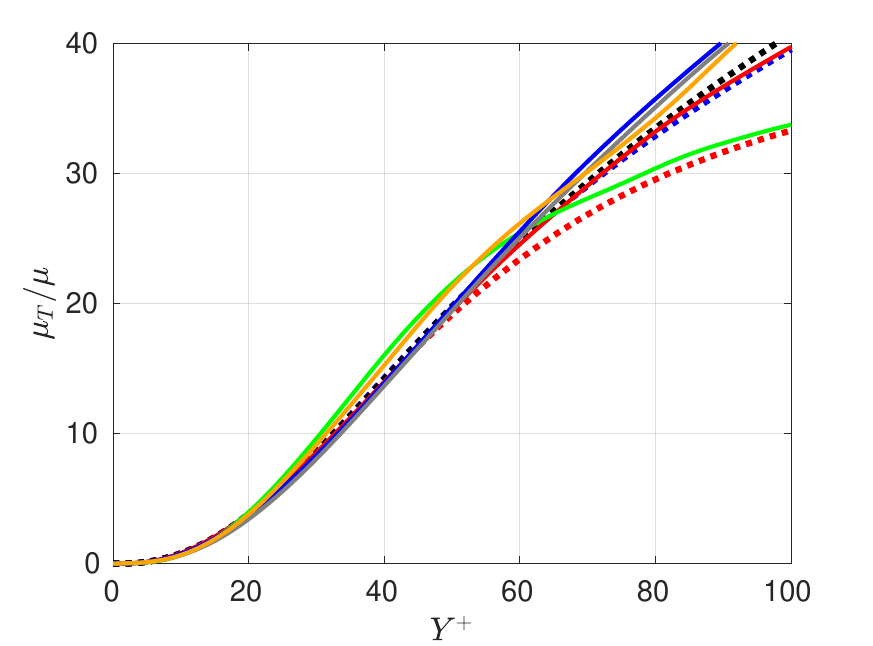}
        \end{subfigure}
        \begin{subfigure}{0.485\textwidth}
            \centering
            \includegraphics[width=\textwidth,trim={3.4cm 8.5cm 3.4cm 8.5cm},clip]{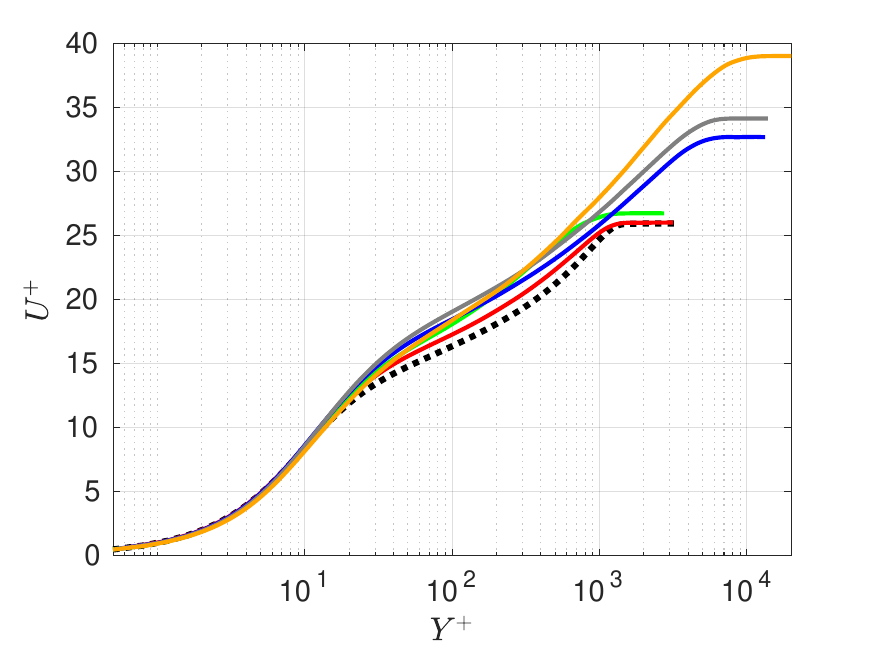}
            \caption{Trettel and Larson \cite{trettel2016}}
        \end{subfigure}
        \begin{subfigure}{0.485\textwidth}
            \centering
            \includegraphics[width=\textwidth,trim={3.4cm 8.5cm 3.4cm 8.5cm},clip]{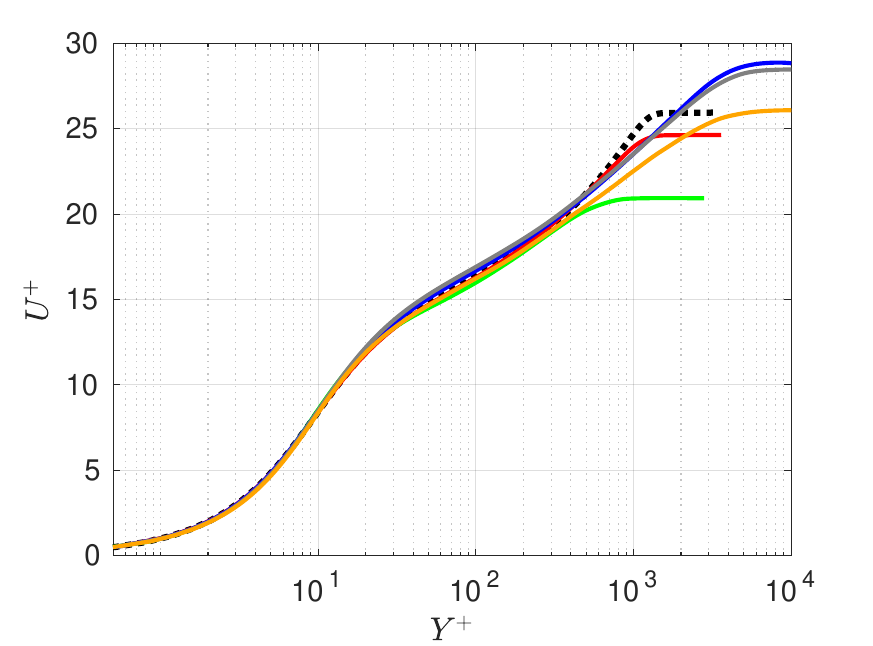}
            \caption{Volpiani et al.\cite{volpiani2020}}
        \end{subfigure}
    \end{center}
    \caption{Nondimensional profiles obtained with different compressibility transformations proposed in \cite{trettel2016,volpiani2020}. Eddy viscosity profiles are on the first row and the transformed velocity profiles are on the second row. Solid lines are compressible DNS \cite{Zhang2018} test cases: \legendline{red}~ M2p5, \legendline{green} M6Tw025, \legendline{blue} M6Tw076, \legendline{gray} M8Tw048, and \legendline{orange} M14Tw018. Symbols are incompressible DNS \cite{schlatter2010} at various Reynolds numbers: \legendsquare{red}~$Re_\tau=492$, \legendsquare{blue}~$Re_\tau=974$, and \legendsquare{black}~$Re_\tau=1271$.}
    \label{fig:profiles-other-methods}
\end{figure}

\begin{figure}[htbp]
    \begin{center}
        \begin{subfigure}{0.485\textwidth}
            \centering
            \includegraphics[width=\textwidth,trim={3.4cm 8.5cm 3.4cm 8.5cm},clip]{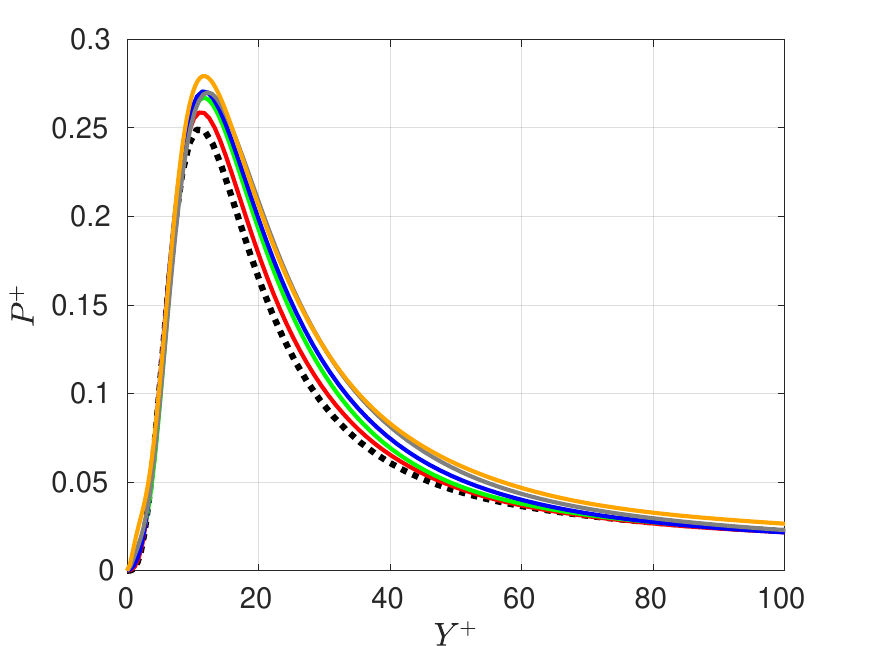}
        \end{subfigure}
        \begin{subfigure}{0.485\textwidth}
            \centering
            \includegraphics[width=\textwidth,trim={3.4cm 8.5cm 3.4cm 8.5cm},clip]{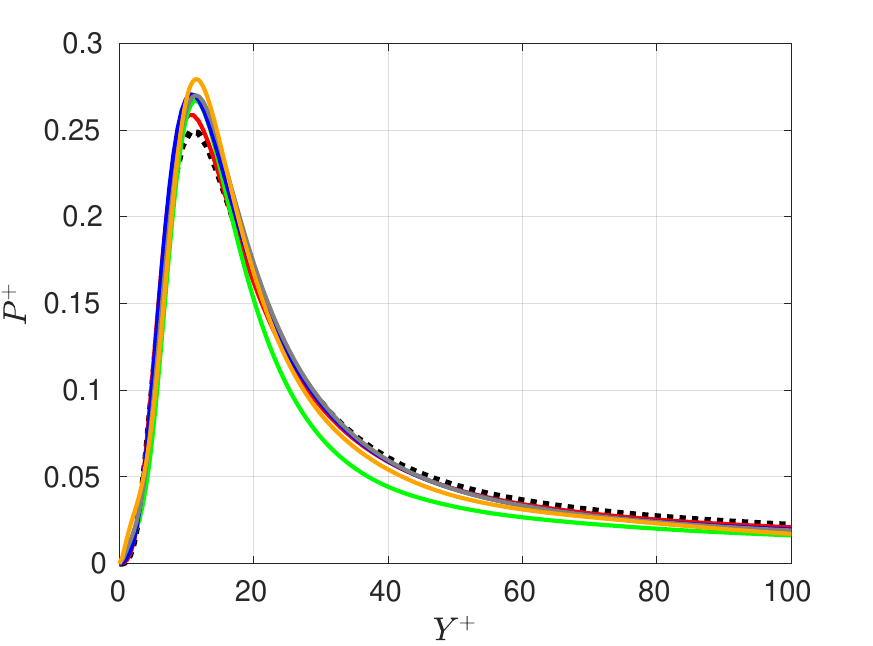}
        \end{subfigure}
    \end{center}
    \caption{Nondimensional profiles of TKE production obtained with different compressibility transformations proposed in \cite{trettel2016,volpiani2020}. Legends are the same as those in \Cref{fig:profiles-other-methods}}
    \label{fig:prod-profiles-other-methods}
\end{figure}

The significance of satisfying eddy viscosity equivalence is demonstrated in \Cref{fig:profiles-other-methods}. The transformation of Trettel and Larsson \cite{trettel2016} simply sets $Y^+=y\sqrt{\rho\tau}/\mu$ from the semi-local scaling and uses an integral formula to compute $U^+$. It is obvious that eddy viscosity profiles do not collapse on incompressible profiles for $Y^+>10$, which generates a mismatch in the y-intercept of the log law corresponding to the transformed velocity profiles. The alternative transformation proposed by Volpiani et al. \cite{volpiani2020} shows a much better collapse of the eddy viscosity profiles. This directly results in a better collapse of transformed velocity onto incompressible profiles. Despite that success, however, eddy viscosity profiles show incorrect sensitivity to the strong wall cooling of M6Tw025 and M14Tw018 cases. This is also reflected in the nondimensional TKE production profiles, as shown in \Cref{fig:prod-profiles-other-methods}. 

In this work, we are mainly focused on the logarithmic layer and below. We propose several accurate transformations for the inner region of wall-bounded hypersonic turbulence. For the sake of completeness, we will also provide the transformed incompressible velocity profiles in the outer region as well, without expecting them to be as accurate. Motivated by this success and inspired by the outer layer scaling approach in \cite{hasan2024}, we finally introduce a transformation that improves the accuracy in the outer region. 

\section{Methods and Results}
In this section, we introduce several new compressibility transformations for computing the incompressible wall-normal coordinate $Y$. All of the proposed formulations aim to satisfy eddy viscosity equivalence for $Y^+\lessapprox40$ and $\partial U/\partial Y=u_\tau/(\kappa Y)$ for a given Von Karman constant $\kappa$ in the logarithmic layer. The transformed, incompressible velocity $U$ is computed after this step by integrating \Cref{eq:vel-grad-hypothesis} with respect to the transformed wall-normal coordinate $Y$.  
\subsection{Method 1}
This transformation is an example designed to illustrate the importance of satisfying the eddy viscosity equivalence below the logarithmic layer. First, we divide the boundary layer into three distinct regions,
\begin{equation}
    Y(y) = 
    \begin{cases}
        Y_1(y), & \text{for } y<y_{log,1}\\
        Y_2(y), & \text{for } y_{log,1}\le y\le y_{log,2}\\
        Y_3(y), & \text{otherwise}
    \end{cases}
\end{equation}
Here, $y_{log,1}$ is a representative lower bound for the logarithmic layer and its estimation will be discussed later. The upper bound for the logarithmic layer is denoted by $y_{log,2}$, and for simplicity, it is determined by $y_{log,2}=0.2\delta_{99}$, where $\delta_{99}$ is the standard boundary layer thickness and available in the compressible DNS data set. 

\begin{figure}
    \centering\includegraphics[width=0.65\textwidth,trim={3.4cm 8.5cm 3.4cm 8.5cm},clip]{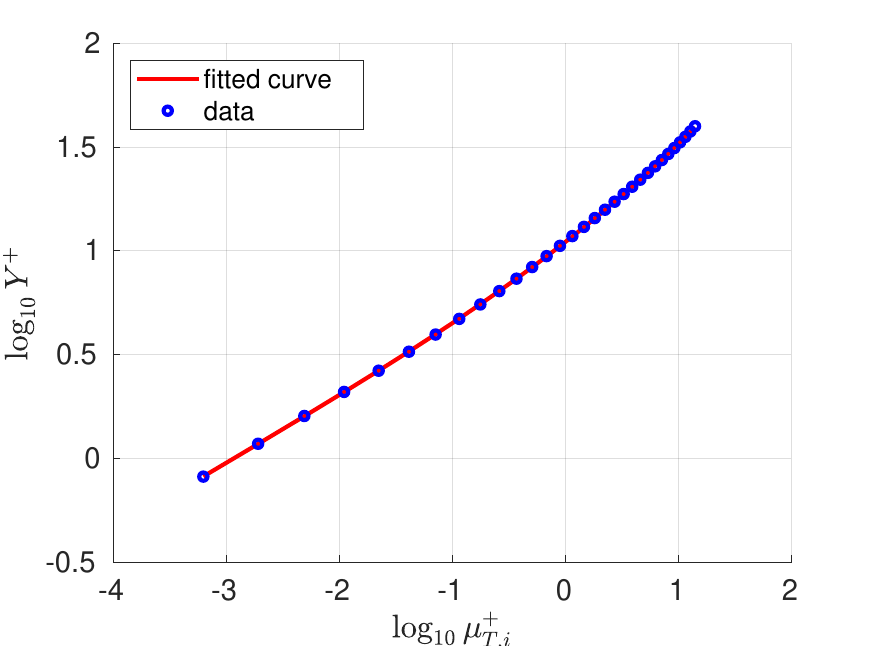}
    \caption{Curve fit for $Y^+$ vs $\mu_{T,i}^+$ from the incompressible data set \cite{schlatter2010} up to $Y^+\lessapprox40$}
    \label{fig:mut-fit}
\end{figure}

To determine $Y_1(y)$ below the logarithmic layer, we follow a straightforward approach and compare $\mu_T^+(y)$ from a compressible DNS to $\mu_{T,i}^+(Y)$ from an incompressible DNS. Since the eddy viscosity equivalence implies $\mu_T^+=\mu_{T,i}^+$, there exists  a one-to-one mapping between $Y_1$ and $y$ such that 
\begin{equation}
    Y_1^+=Y_1^+(\mu_{T,i}^+)=Y_1^+(\mu_{T}^+(y))=Y_1^+(y)
\end{equation}
This one-to-one mapping can be obtained by considering a curve fit in the form of $Y^+=Y^+(\mu_{T,i}^+)$ from an available incompressible DNS data set. For example, the curve fit
\begin{equation}\label{eq:mut-fit}
    \log_{10}{Y_1^+}= 4.7896\,x^4+ 7.1834\,x^3+ 4.0672\,x^2+ 4.25\,x+ 1.0417
\end{equation}
is shown in \Cref{fig:mut-fit} for the data set in \cite{schlatter2010} for $Y^+\lessapprox40$ with at least 95\% confidence, where $x=\left(\log_{10}{\mu_{T,i}^+}\right)/10$.  The compressible DNS provide $\mu_T(y)=-\overline{\rho u^{''}v^{''}}/(\partial u/\partial y)$.  Following  \Cref{eq:mut-equivalence},  this was substituted into \Cref{eq:mut-fit} to give $Y_1(y)$ directly.

In the logarithmic region, we simply assume that the incompressible velocity profile follows the log-law
\begin{equation}
     \frac{\partial U}{\partial Y}=\frac{u_\tau}{\kappa Y(y)}
\end{equation}
Using the velocity gradient assumption in \Cref{eq:vel-grad-hypothesis}, an analytical relation for $Y(y)$ is obtained
\begin{equation}\label{eq:log-Y-y-formula}
    Y_2(y) = \frac{u_\tau}{\kappa}\left(\frac{\mu_w}{\mu}\right)\left(\frac{\partial u}{\partial y}\right)^{-1}
\end{equation}

For the outer-region, a reasonable velocity transformation is obtained by setting 
\begin{equation}
    Y_3(y)=\left(\frac{\mu_w}{\mu}\right)\left(\frac{\rho}{\rho_w}\right)^{1/2}y,
\end{equation}
Note that this is equivalent to $Y_3^+=y^*$, where $y^*$ is the wall-normal coordinate normalized by the semi-local scaling.

The formulation of Method 1 is completed by approximating the lower bound of the logarithmic layer as
\begin{equation}
    y_{log,1}=\argmin_{10\le Y_1^+\le 50}{\left|Y_1(y)-Y_2(y)\right|}
\end{equation}

\begin{figure}[htbp]
    \begin{center}
        \begin{subfigure}{0.485\textwidth}
            \centering
            \includegraphics[width=\textwidth,trim={3.4cm 8.5cm 3.4cm 8.5cm},clip]{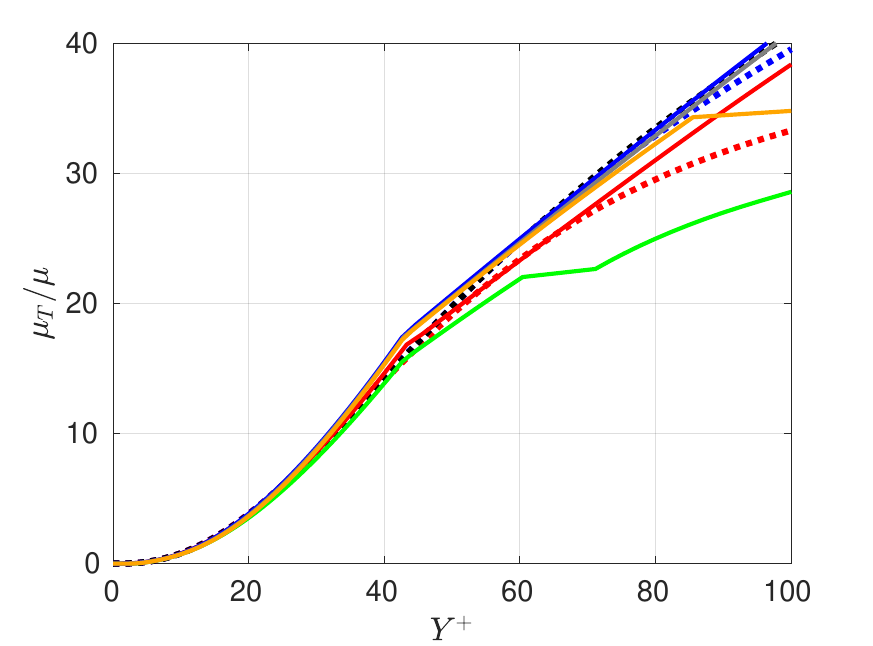}
            \caption{Eddy Viscosity}
        \end{subfigure}
        \begin{subfigure}{0.485\textwidth}
            \centering
            \includegraphics[width=\textwidth,trim={3.4cm 8.5cm 3.4cm 8.5cm},clip]{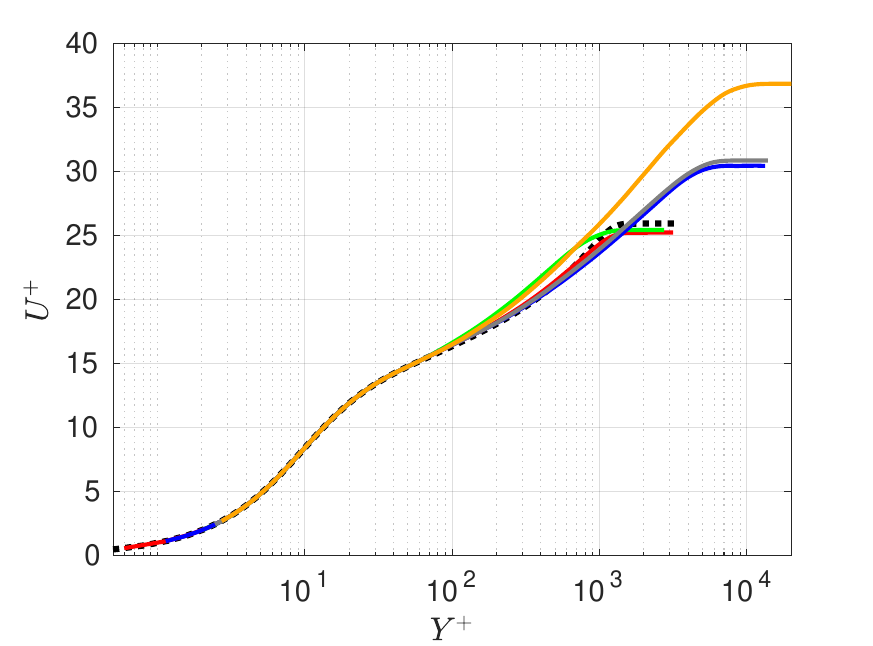}
            \caption{Velocity}
        \end{subfigure}
        \begin{subfigure}{0.485\textwidth}
            \centering
            \includegraphics[width=\textwidth,trim={3.3cm 8.5cm 3.3cm 8.5cm},clip]{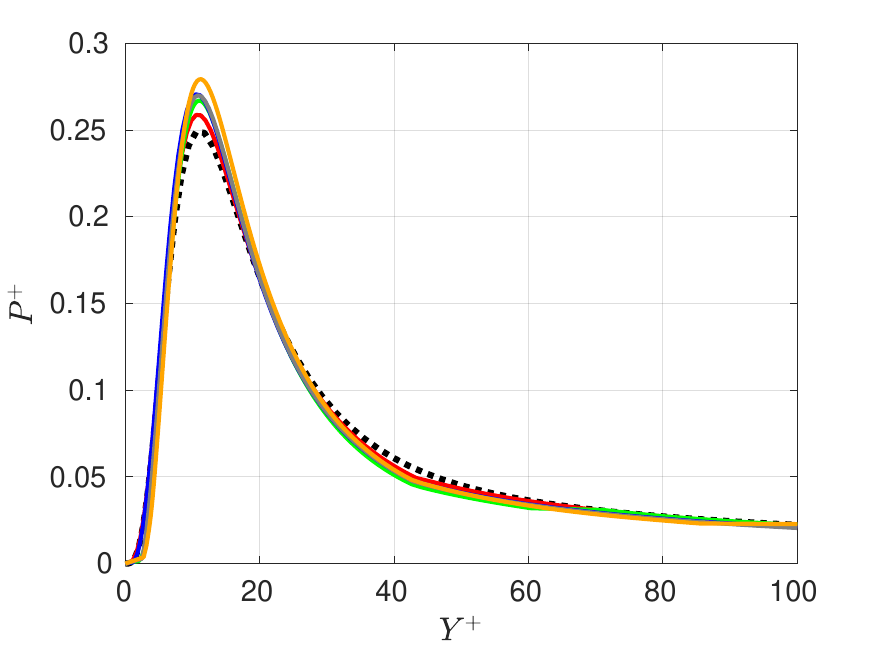}
            \caption{TKE Production}\label{fig:mtd-1c}
        \end{subfigure}
        \begin{subfigure}{0.485\textwidth}
            \centering
            \includegraphics[width=\textwidth,trim={3.3cm 8.5cm 3.3cm 8.5cm},clip]{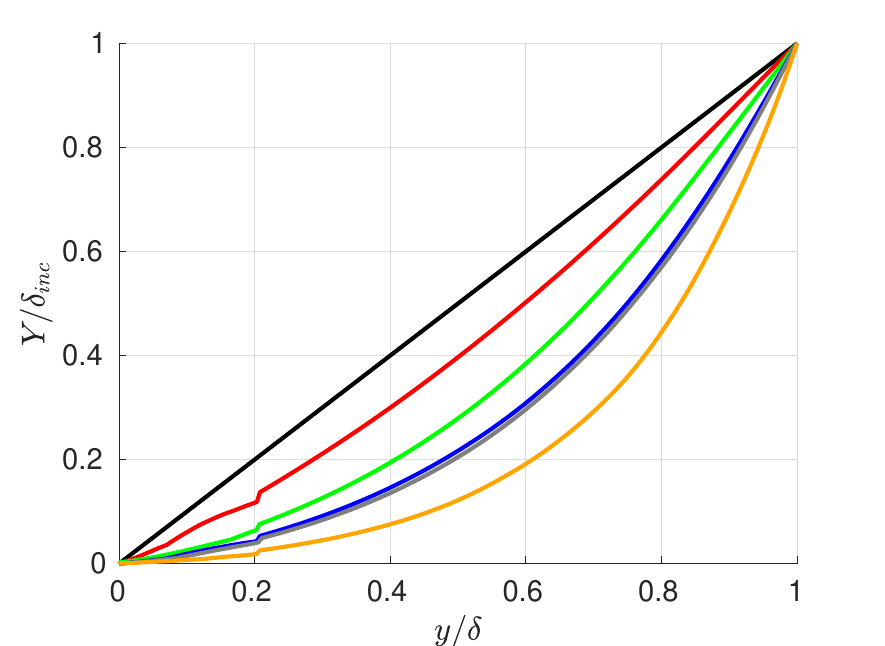}
            \caption{Transformed wall-normal coordinate}\label{fig:mtd-1d}
        \end{subfigure}
    \end{center}
    \caption{Transformed Profiles with Method 1. Solid black line is for $Y/\delta_{inc}=y/\delta$ and other legends are the same as those in \Cref{fig:profiles-other-methods}.}
    \label{fig:mtd-1}
\end{figure}

\Cref{fig:mtd-1} shows the transformed profiles of eddy viscosity, velocity and TKE production. Below the logarithmic layer, the eddy viscosity profiles collapse onto the incompressible profiles. There are kinks while switching from $Y_2$ to $Y_3$, but this is the region where the eddy viscosity equivalence is no longer sought. Recall that the intent of this example is to illustrate the effect of satisfying the eddy viscosity scaling, which is obvious in the transformed velocity profiles in  most of the logarithmic layer and below. The compressible profiles are almost indistinguishable as they collapse perfectly onto the incompressible velocity profile in this region. This success is reflected in the relative errors of the transformed velocity, as reported in \Cref{tab:rel-err-mtd-1}. While the transformation presented in Volpiani et al. \cite{volpiani2020} enjoys an accuracy with errors less than 3.5 \%, errors of Method 1 consistently remains less than 0.65 \%. \Cref{fig:mtd-1c} also shows a remarkable success in collapsing the compressible TKE production profiles on the incompressible profiles for $Y^+\le100$, which is the best performance reported in the literature so far, to the best of our knowledge. The transformed wall-normal coordinates are compared in \Cref{fig:mtd-1d}. In the adiabatic supersonic case M2p5, the transformed coordinate is below the incompressible line, suggesting that higher Mach numbers push the transformed coordinates further below the incompressible line. A comparison of M6Tw025 and M6Tw076 suggests an opposite effect of wall cooling, as M6Tw025 is closer to the incompressible line. This is supported by the collapse of M6Tw076 onto M8Tw048. In that, high Mach number effects may have been balanced by higher cooling rates in M8Tw048 when collapsing on M6Tw076. The opposite effects of increasing Mach numbers and wall cooling has long been recognized in the literature \cite{huang1994}. In this case, it might be argued that Mach number effects in M14Tw018 play a more prominent role for the transformation than the those of the strong wall cooling, as $Y(y)$ profile is further below M6Tw076 and M8Tw048.

\begin{table}[htbp]
    \centering
    \begin{tabular}{lccc}\hlineB{3}
        Case     & Trettel and Larsson \cite{trettel2016} & Volpiani et al. \cite{volpiani2020} & Present Study (Method 1) \\\hline
        M2p5     &          5.61                          &    0.83                             &   0.65                   \\
        M6Tw025  &          9.60                          &    1.84                             &   0.14                   \\ 
        M6Tw076  &          11.93                         &    2.07                             &   0.65                   \\
        M8Tw048  &          15.95                         &    3.51                             &   0.65                   \\
        M14Tw018 &          12.86                         &    3.16                             &   0.48                   \\\hlineB{3}
    \end{tabular}
    \caption{Relative errors (percent) of transformed velocity profiles in the inner layer for $y\le0.2\delta_{99}$.}
    \label{tab:rel-err-mtd-1}
\end{table}

\subsection{Method 2}
This method considers a transformation of the form:
\begin{equation}
    Y(y) = 
    \begin{cases}
        Y_1(y), & \text{for } y<0.2\delta_{99}\\
        Y_2(y), & \text{otherwise}
    \end{cases}
\end{equation}

Instead of curve fitting as in the previous method, this method employs an incompressible mixing length model and Van Driest damping, to satisfy the eddy viscosity hypothesis and log-law in the inner layer for $y<0.2\delta_{99}$. The incompressible mixing length model is defined as 
\begin{equation}
    \mu_{T,i}^+=\frac{\rho_w}{\mu_w}\ell_m^2\frac{\partial U}{\partial Y},
\end{equation}
where $\ell_m$ is the mixing length. Using the velocity gradient hypothesis \Cref{eq:vel-grad-hypothesis}, the mixing length model becomes
\begin{equation}\label{eq:mut-mixing-length}
    \mu_{T,i}^+=\frac{\rho_w\mu}{\mu_w^2}\ell_m^2\frac{\partial u}{\partial y}.
\end{equation}
Next, we invoke the Van Driest damping for an incompressible flow
\begin{equation}\label{eq:ellm}
    \ell_m = \frac{\mu_w}{\rho_wu_\tau}\kappa Y_1^+\left(1-\exp{\left(-\frac{Y_1^+}{A^+}\right)}\right),
\end{equation}
and set $\kappa=0.41$ and $A^+=26$. Rewriting the eddy viscosity equivalence in the form $\mu_T^+-\mu_{T,i}^+=0$ and substituting \Cref{eq:ellm} in \Cref{eq:mut-mixing-length} 
\begin{equation}\label{eq:residual-mixing-length}
    \mu_{T}^+-\left(\frac{\mu}{\rho_wu_\tau^2}\right)\left[\kappa Y_1^+\left(1-\exp{\left(-\frac{Y_1^+}{A^+}\right)}\right)\right]^2\frac{\partial u}{\partial y} = 0. 
\end{equation}
The premise of scaling is that $u(y)$ is given, and proper scaling should collapse it onto
$U(Y)$, via \Cref{eq:vel-grad-hypothesis}: hence, $\rho,\mu,\partial u/\partial y$ are known from the DNS data set and substituting
\begin{equation}
    \mu_T^+=\frac{\tau_w}{\mu\partial u/\partial y}-1
\end{equation}
in \Cref{eq:residual-mixing-length} provides a closed equation  for the transformation $Y_1(y)$.
This makes \Cref{eq:residual-mixing-length}  a nonlinear equation for $Y_1$, which can easily be solved using Newton's method. Finally, Method 2 is completed by setting $Y_2$ using the semi-local scaled $y^*$ as in the previous example in the outer region for $y\ge0.2\delta_{99}$.
  
\begin{figure}[htbp]
    \begin{center}
        \begin{subfigure}{0.485\textwidth}
            \centering
            \includegraphics[width=\textwidth,trim={3.4cm 8.5cm 3.4cm 8.5cm},clip]{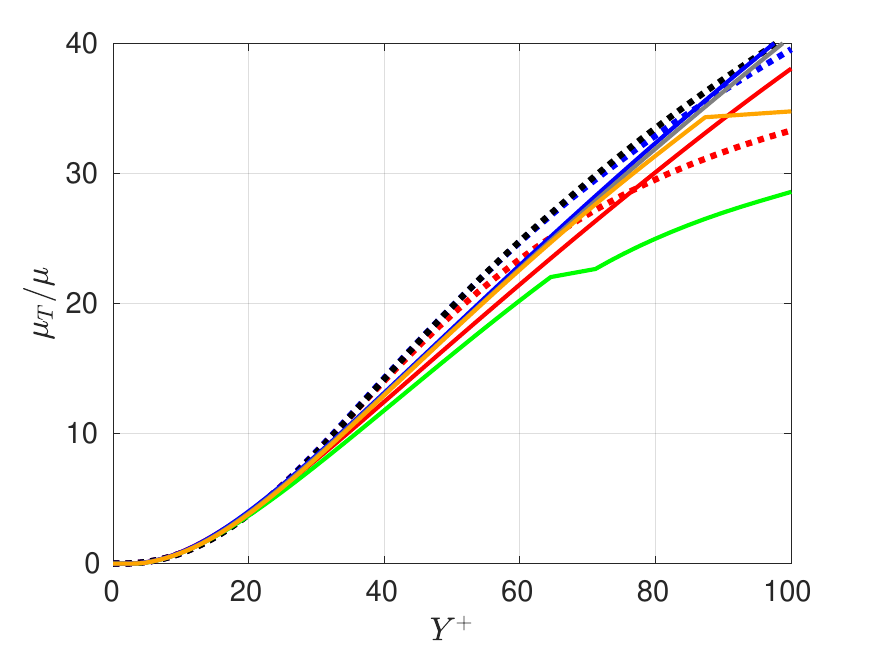}
            \caption{Eddy Viscosity}
        \end{subfigure}
        \begin{subfigure}{0.485\textwidth}
            \centering
            \includegraphics[width=\textwidth,trim={3.4cm 8.5cm 3.4cm 8.5cm},clip]{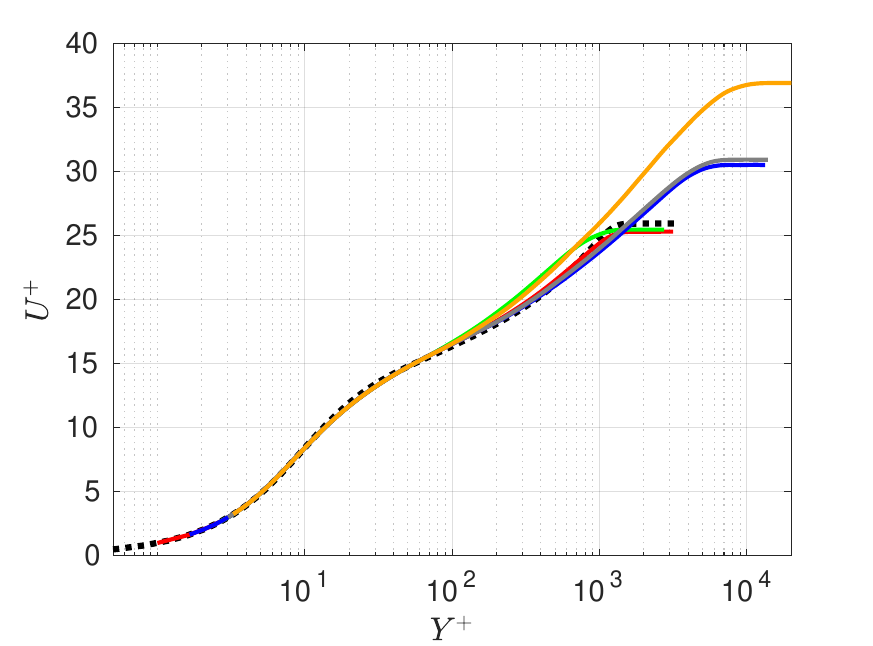}
            \caption{Velocity}
        \end{subfigure}
        \begin{subfigure}{0.485\textwidth}
            \centering
            \includegraphics[width=\textwidth,trim={3.3cm 8.5cm 3.3cm 8.5cm},clip]{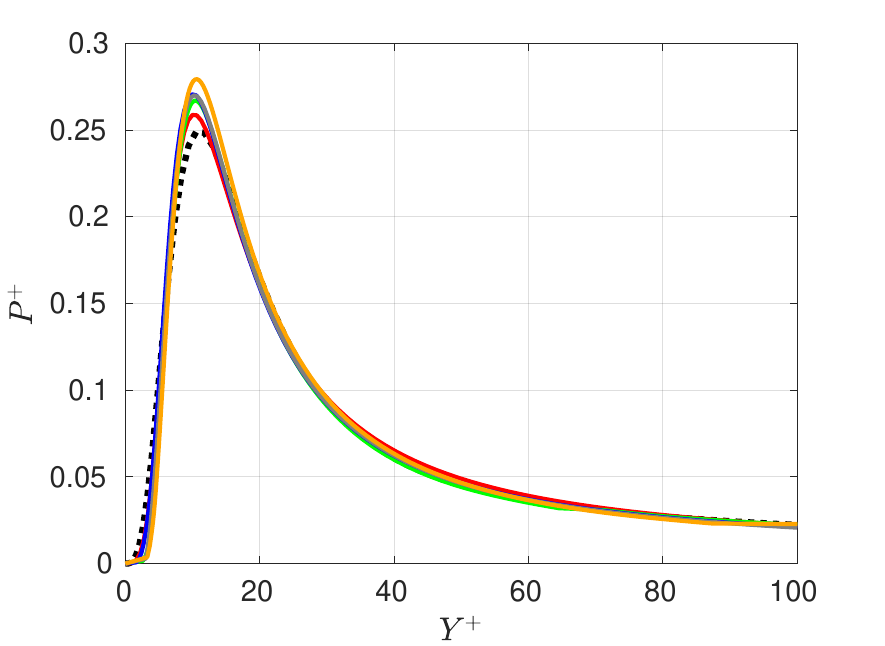}
            \caption{TKE Production}\label{fig:mtd-2c}
        \end{subfigure}
        \begin{subfigure}{0.485\textwidth}
            \centering
            \includegraphics[width=\textwidth,trim={3.3cm 8.5cm 3.3cm 8.5cm},clip]{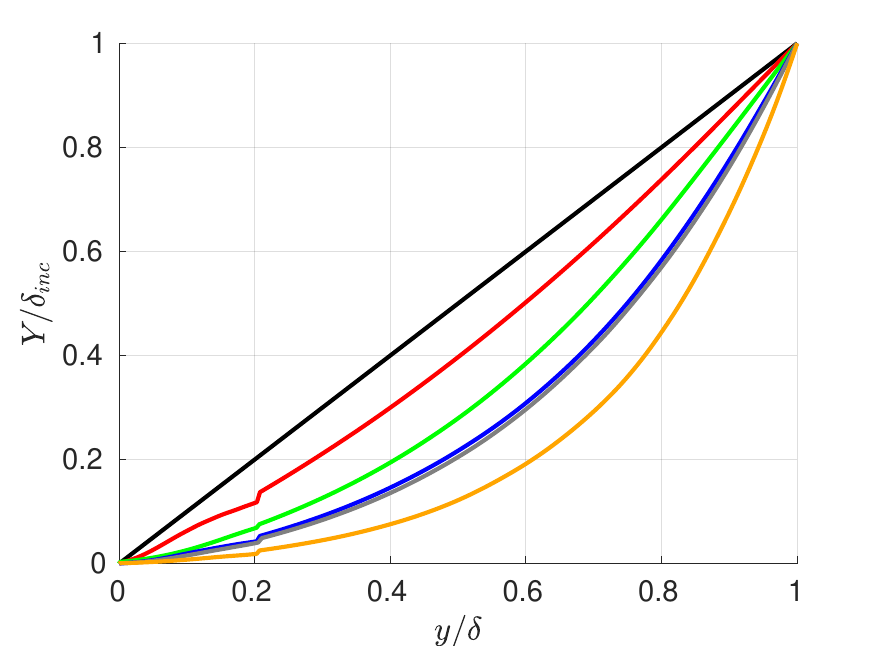}
            \caption{Transformed wall-normal coordinate}\label{fig:mtd-2d}
        \end{subfigure}
    \end{center}
    \caption{Transformed Profiles with Method 2. Legends are the same as those in \Cref{fig:mtd-1}.}
    \label{fig:mtd-2}
\end{figure}

\Cref{fig:mtd-2} shows the transformed profiles. The present transformed compressible eddy viscosity profiles matches the incompressible profiles for $Y^+\lessapprox40$  with a reasonable agreement for all Mach numbers and it does not show sensitivity to wall cooling effects like the transformation of Volpiani et al. \cite{volpiani2020}. It is worth emphasizing that we used the incompressible Van Driest damping parameters without any additional compressibility treatments in this region. The performance of the transformed velocity is slightly worse than Method 1 with a maximum relative error of $1.93\,\%$ in the inner layer, which is still better than Volpiani et al. \cite{volpiani2020}. Recall that $Y$ in the inner region --including the logarithmic layer-- was computed by a nonlinear Van Driest damping formulation. This approach produced excellent collapse of compressible TKE production profiles to incompressible profiles as shown in \Cref{fig:mtd-2c}. Similar to the previous Method, \Cref{fig:mtd-2d} shows a similar behavior of the transform wall-normal coordinate profiles.

\subsection{Method 3 (modified Volpiani transformation)}
The previous examples show the significance of satisfying the eddy viscosity equivalence below the logarithmic layer. This gives rise to the question whether existing transformations can be improved by integrating them with this hypothesis. In this example, we present  a modified Volpiani transformation. It uses the incompressible mixing length model and Van Driest damping below the logarithmic layer as described in Method 2. Above this region, it sets $Y$ from the transformation of Volpiani et al. \cite{volpiani2020}:
\begin{equation}
    Y(y) = 
    \begin{cases}
        Y_1(y), & \text{for } y<y_{log,1}\\
        Y_2(y), & \text{otherwise}
    \end{cases}
\end{equation}
Here, $y_{log,1}$ is computed as in Method 1 and the transformation of Volpiani et al. \cite{volpiani2020}  is modified as
\begin{equation}
    Y_2(y) = Y_1(y_{log,1})+\int_{y_{log,1}}^{y} \left(\frac{\rho}{\rho_w}\right)^{1/2}\left(\frac{\mu_w}{\mu}\right)^{3/2}\;dy'
\end{equation}
Note that the lower bound of this transformation is different from that in \cite{volpiani2020}. In that, it is changed from $y=0$ to $y=y_{log,1}$ in Method 3. Note also that, compared to previous examples, this transformation does not use $y^*$ to set the transformed coordinate in the outer layer.

\begin{figure}[htbp]
    \begin{center}
        \begin{subfigure}{0.485\textwidth}
            \centering
            \includegraphics[width=\textwidth,trim={3.4cm 8.5cm 3.4cm 8.5cm},clip]{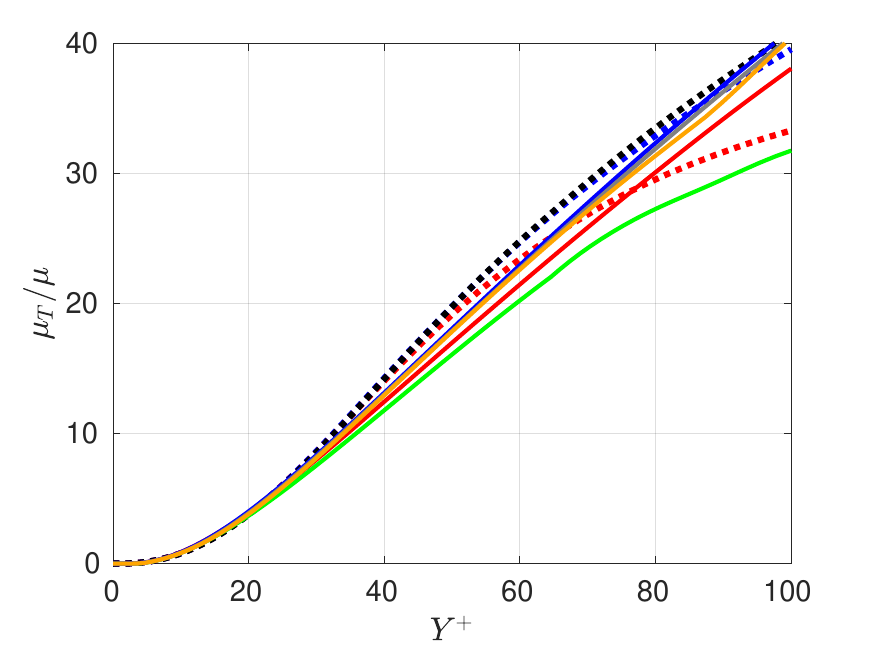}
            \caption{Eddy Viscosity}
        \end{subfigure}
        \begin{subfigure}{0.485\textwidth}
            \centering
            \includegraphics[width=\textwidth,trim={3.4cm 8.5cm 3.4cm 8.5cm},clip]{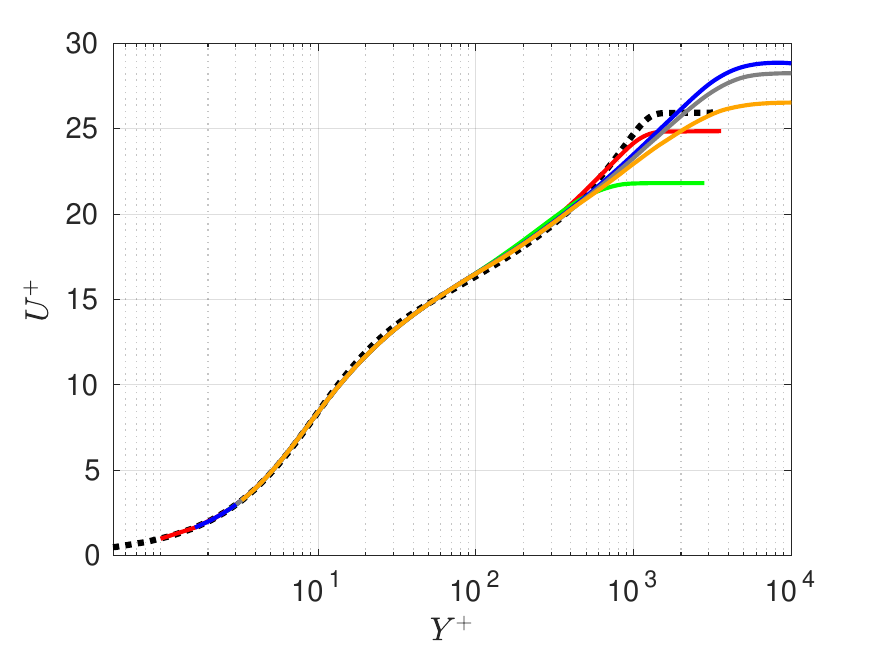}
            \caption{Velocity}
        \end{subfigure}
        \begin{subfigure}{0.485\textwidth}
            \centering
            \includegraphics[width=\textwidth,trim={3.3cm 8.5cm 3.3cm 8.5cm},clip]{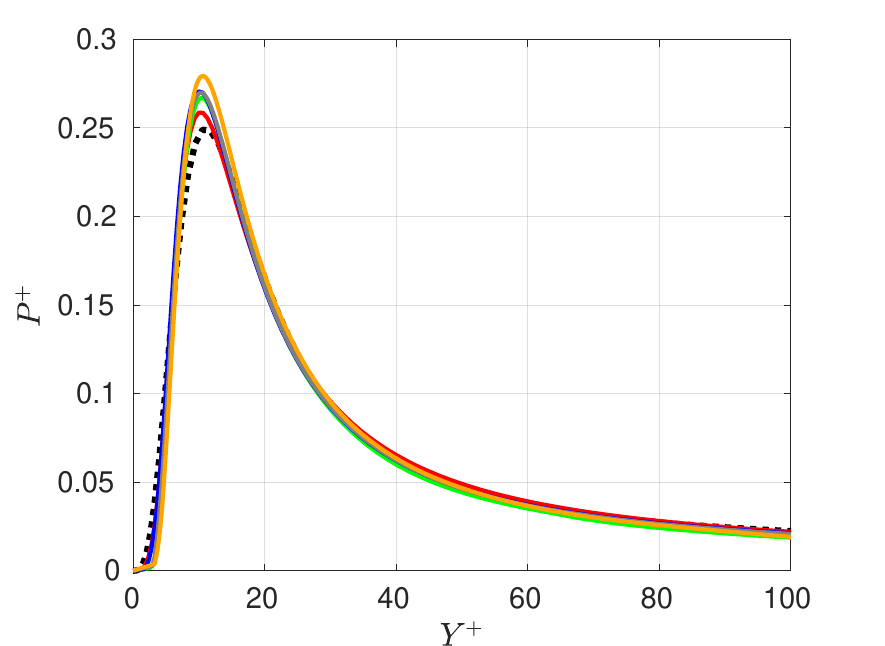}
            \caption{TKE Production}\label{fig:mtd-3c}
        \end{subfigure}
        \begin{subfigure}{0.485\textwidth}
            \centering
            \includegraphics[width=\textwidth,trim={3.3cm 8.5cm 3.3cm 8.5cm},clip]{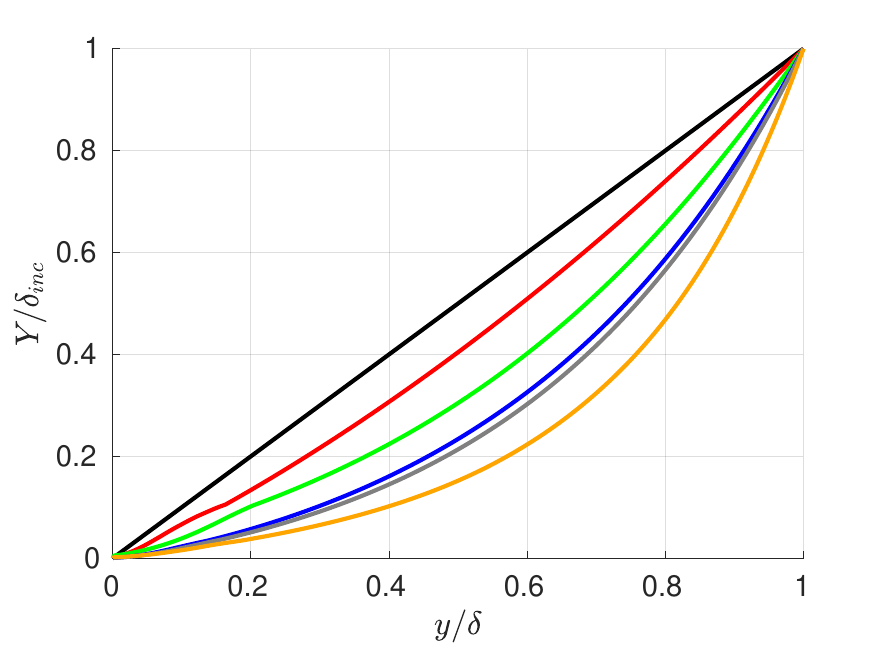}
            \caption{Transformed wall-normal coordinate}\label{fig:mtd-3d}
        \end{subfigure}
    \end{center}
    \caption{Transformed Profiles with Method 3. Legends are the same as those in \Cref{fig:mtd-1}.}
    \label{fig:mtd-3}
\end{figure}

The improved performance of the modified Volpiani transformation is shown in \Cref{fig:mtd-3}. The overall performance is quite similar to Method 2 in the inner layer. The compressible Van Driest treatment removed the dependence on cold wall effects for the eddy viscosity profiles and collapsed them onto the incompressible profiles below the logarithmic layer. This resulted in a decrease of maximum relative error in velocity for the Volpiani transformation from $3.5\,\%$ to $1.9\,\%$ in the inner layer. The accuracy can be improved further by using Method 1 below the logarithmic layer. In that case, the maximum error is decreased to $0.83\,\%$. As in the previous examples, the transform wall-normal coordinate profiles are similar and the transformed TKE production does not display a significant dependence on the Mach number and wall cooling as these profiles are collapsed on the incompressible profiles for $Y^+\le100$.

\subsection{Method 4 (Reynolds number-based transformation-A)}
In this example, we propose a new hypothesis to derive a new compressibility transformation from a claim that, for a compressible Reynolds number definition, there should exist an equivalent incompressible Reynolds number, e.g. $Re_{comp}=Re_{incomp}$, under a suitable compressibility transformation $y\rightarrow Y$. This assumption is mainly motivated by Morkovin's hypothesis,  that the dynamics of turbulence is largely a function of the changes in the fluid properties. As a secondary motivation, our considerations can be very useful for deriving a consistent Reynolds number definition for the analysis of hypersonic turbulent boundary layers. Numerous ad-hoc attempts have made to relate compressible boundary layers to incompressible boundary layers in the hopes of finding a proper Reynolds number definition: most prominent examples are $Re_\tau$, $Re_\tau^*$, $Re_\delta$ and $Re_{\delta2}$ (see \cite{Zhang2018} for their definitions).  

We choose
\begin{equation}\label{eq:Re-incomp}
    Re_{inc}(Y)\equiv\int_{0}^{Y}\frac{\rho_w u_\tau}{\mu_w}dY'=\frac{\rho_w u_\tau Y}{\mu_w}.
\end{equation}
Note that at $Y=\delta_i$, where $\delta_i$ is the boundary layer thickness of the corresponding incompressible flow, $Re_{inc}(\delta_i)$ reduces to the traditional $Re_\tau=\rho_w u_\tau \delta_i/\mu_w$.

In the same spirit, the compressible Reynolds number can be defined as
\begin{equation}\label{eq:Re-comp}
    Re_{comp}(y)=\int_{0}^{y}\frac{\rho u_c}{\mu}dy'
\end{equation}
where $u_c$ is a compressible velocity scale (possibly a function of $y$) to be determined. Since $\rho\rightarrow\rho_w$ and $\mu\rightarrow\mu_w$ when the compressibility and wall cooling effects disappear as $M\rightarrow0$ and $T_r\rightarrow T_w$, $u_c$ needs to satisfy $u_c\rightarrow u_\tau$ as a consistency condition.

The Reynolds number equivalence suggests a relation between $y$ and $Y$. By the fundamental theorem of Calculus, we obtain
\begin{equation}\label{eq:y-Y-relation}
    \frac{dY}{dy}=\left(\frac{\rho}{\rho_w}\right)\left(\frac{\mu_w}{\mu}\right)\left(\frac{u_c}{u_\tau}\right)
\end{equation}
Note that this is a generalization of the Volpiani transformation \cite{volpiani2020}, which can be easily seen by setting $u_c=\sqrt{(\rho_w/\rho)(\mu_w/\mu)}$.

To derive an analytical expression for $u_c$ in the logarithmic region, we assume the incompressible velocity profile follows the log-law. Differentiating \Cref{eq:log-Y-y-formula} with respect to y and substituting it into \Cref{eq:y-Y-relation} give
\begin{equation}\label{eq:u_c-log-layer}
    \frac{u_c}{u_\tau} = -\frac{1}{\kappa}\frac{\rho_w}{\rho}\frac{\frac{\partial}{\partial y}\left(\mu\frac{\partial u}{\partial y}\right)}{\mu\left(\frac{\partial u}{\partial y}\right)^2} u_\tau
\end{equation}
As $M\rightarrow 0$ and $T_r\rightarrow T_w$, we have that $\rho\rightarrow\rho_w$, $\mu\rightarrow\mu_w$, $y\rightarrow Y$, and $\partial_y u\rightarrow u_\tau/\kappa Y$. Then, 
\begin{equation}
    \frac{u_c}{u_\tau}\rightarrow -\frac{1}{\kappa}\frac{\rho_w}{\rho_w}\frac{\frac{\partial}{\partial Y}\left(\mu_w\frac{u_\tau}{\kappa Y}\right)}{\mu_w\left(\frac{u_\tau}{\kappa Y}\right)^2} u_\tau=1,
\end{equation}
which shows consistency in the incompressible limit.

Note that this derivation holds only for the logarithmic layer. Estimating the region below the logarithmic layer is left to the next example and we set $Y$ using the curve fitting approach described in Method 1. In the outer region, we set $u_c=u_\tau$ for simplicity. In summary, Method 4 uses
\begin{equation}
    Y(y) = 
    \begin{cases}
        Y_1(y), & \text{for } y<y_{log,1}\\
        Y_2(y), & \text{for } y_{log,1}\le y\le y_{log,2}\\
        Y_3(y), & \text{otherwise},
    \end{cases}
\end{equation}
where $Y_1$ is calculated as Method 1 while 
\begin{equation}
    \begin{aligned}
        Y_2(y) &= Y_1(y_{log,1})+\int_{y_{log,1}}^y\left(\frac{\rho}{\rho_w}\right)\left(\frac{\mu_w}{\mu}\right)\left(\frac{u_c}{u_\tau}\right)\;dy'\\
        Y_3(y) &= Y_2(y_{log,2})+\int_{y_{log,2}}^y\left(\frac{\rho}{\rho_w}\right)\left(\frac{\mu_w}{\mu}\right)\;dy'
    \end{aligned}
\end{equation}

\begin{figure}[htbp]
    \begin{center}
        \begin{subfigure}{0.485\textwidth}
            \centering
            \includegraphics[width=\textwidth,trim={3.4cm 8.5cm 3.4cm 8.5cm},clip]{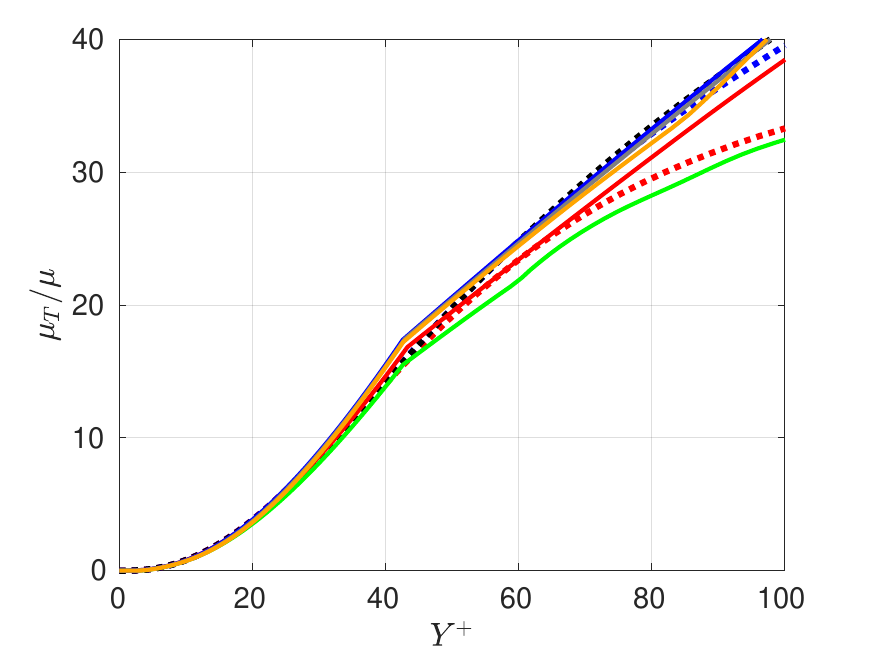}
            \caption{Eddy Viscosity}
        \end{subfigure}
        \begin{subfigure}{0.485\textwidth}
            \centering
            \includegraphics[width=\textwidth,trim={3.4cm 8.5cm 3.4cm 8.5cm},clip]{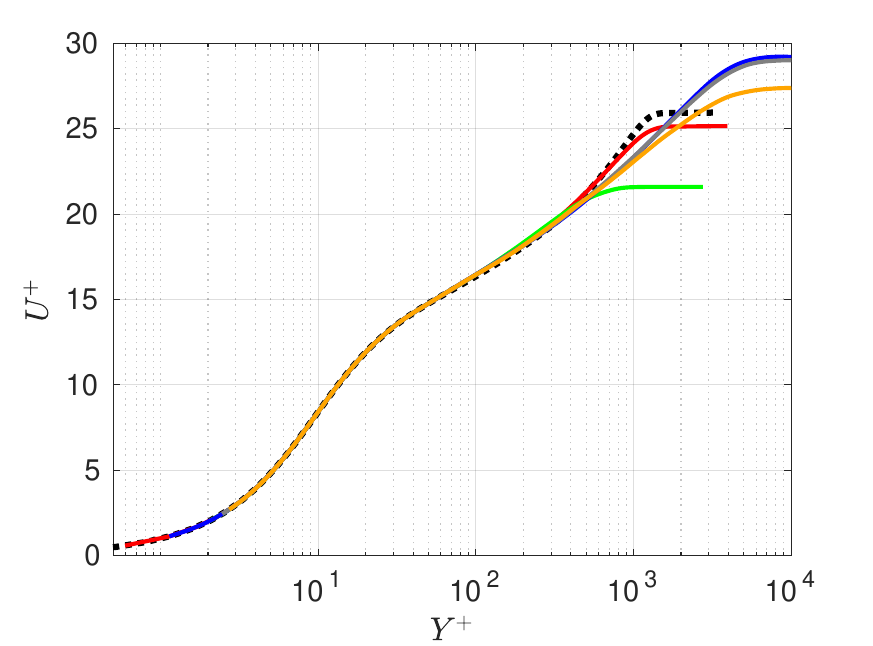}
            \caption{Velocity}
        \end{subfigure}
        \begin{subfigure}{0.485\textwidth}
            \centering
            \includegraphics[width=\textwidth,trim={3.3cm 8.5cm 3.3cm 8.5cm},clip]{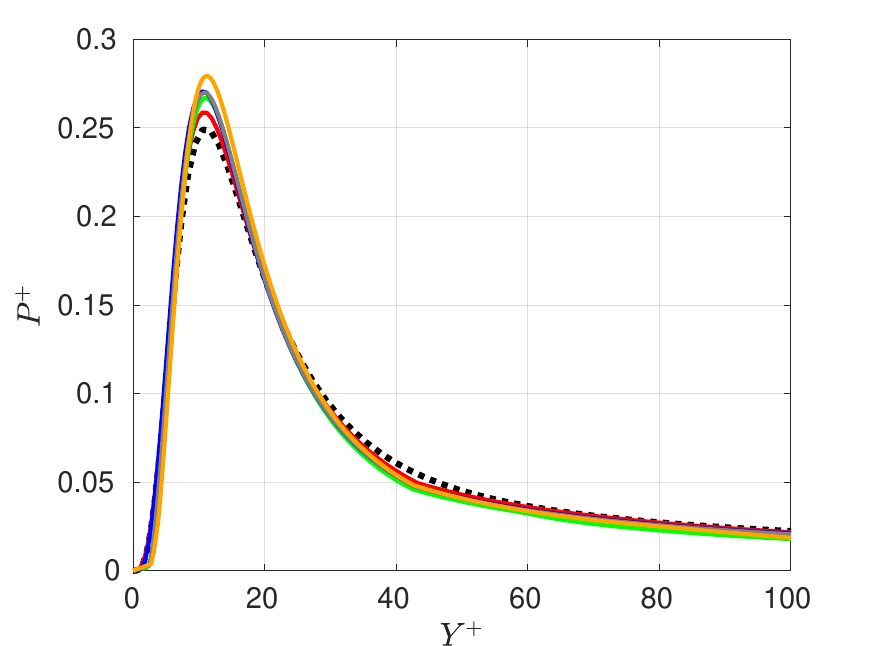}
            \caption{TKE Production}\label{fig:mtd-4c}
        \end{subfigure}
        \begin{subfigure}{0.485\textwidth}
            \centering
            \includegraphics[width=\textwidth,trim={3.3cm 8.5cm 3.3cm 8.5cm},clip]{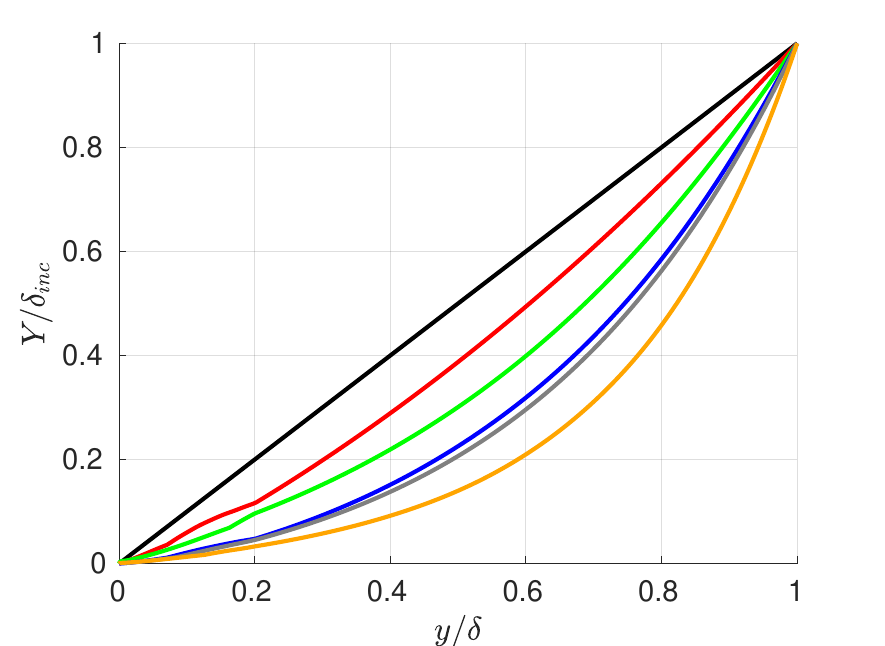}
            \caption{Transformed wall-normal coordinate}\label{fig:mtd-4d}
        \end{subfigure}
    \end{center}
    \caption{Transformed Profiles with Method 4. Legends are the same as those in \Cref{fig:mtd-1}.}
    \label{fig:mtd-4}
\end{figure}

\Cref{fig:mtd-4} shows that Method 4 collapses the compressible eddy viscosity profiles onto incompressible profiles for $Y^+$ more accurately compared to previous examples. Since the curve fitting is used to match the eddy viscosity for $Y^+\lessapprox40$, the transformed velocity profiles have a maximum error of $0.69\%$ in the inner layer for all cases in the compressible DNS data set while the performance in collapsing the TKE production and behavior of the transform wall-normal coordinate profiles are similar to previous examples. Using the Reynolds number equivalence hypothesis, the transformed Reynolds numbers are calculated in \Cref{tab:re-compare}. It is remarkable that the transformed velocity profile of M2p5 Case  collapses nicely onto the incompressible DNS profile, while the transformed Reynolds number with $Re_{\tau,incomp}=1197$ is very close to $Re_{\tau,incomp}=1271$ of the incompressible DNS data. However, the present study should be considered as accurate only for inner layer transformations. The transformed Reynolds number may not be as accurate in other contexts.  

\begin{table}[htbp]
    \centering
    \begin{tabular}{lccc}\hlineB{3}
        Case     & $Re_\tau$ \cite{Zhang2018} & $Re_{\tau,incomp}$ \\\hline
        M2p5     &    510                     &     1197           \\
        M6Tw025  &    450                     &     618            \\ 
        M6Tw076  &    453                     &     3756           \\
        M8Tw048  &    480                     &     3474           \\
        M14Tw018 &    646                     &     2540           \\\hlineB{3}
    \end{tabular}
    \caption{Comparison of compressible and transformed Reynolds numbers for Method 4}
    \label{tab:re-compare}
\end{table}

\subsection{Method 5 (Reynolds number-based transformation-B)}
In this example, we will explore certain choices for $u_c$ for a transformation of the form:
\begin{equation}
    Y(y) = 
    \begin{cases}
        Y_1(y), & \text{for } y<y_{log,2}\\
        Y_2(y), & \text{otherwise},
    \end{cases}
\end{equation}
where
\begin{equation}
    \begin{aligned}
        Y_1(y) &= \int_0^y\left(\frac{\rho}{\rho_w}\right)\left(\frac{\mu_w}{\mu}\right)\left(\frac{u_c}{u_\tau}\right)\;dy'\\
        Y_2(y) &= Y_1(y_{log,2})+\int_{y_{log,2}}^y\left(\frac{\rho}{\rho_w}\right)\left(\frac{\mu_w}{\mu}\right)\;dy'
    \end{aligned}
\end{equation}
Clearly, a natural choice for the compressible velocity scale $u_c$ can be $u_c=u_\tau$, assuming that the compressible velocity scale is the same the incompressible friction velocity and the changes in the Reynolds numbers are due to the changes in the fluid properties. Another reasonable choice is to set $u_c=u_\tau^*$ where $u_\tau^*=\sqrt{\tau_w/\rho}$ is the semi-local friction velocity. In order to include wall heat transfer, one might set $u_c=u_\tau^*+u_q$, where
\begin{equation}
    u_q=\frac{q_w}{\rho C_p T}, \quad q_w =-\left(k\frac{\partial T}{\partial y}\right)_{y=0} 
\end{equation}
and $q_w$ is the heat flux at the wall. A summary of $u_c$ choices is provided in \Cref{tab:mtd-5-summary}.
\begin{table}[htbp]
    \centering
    \begin{tabular}{ccc}              \hlineB{3}
        Method  & $u_c$           & Maximum velocity error ($\%$)\\ \hline
        5a      &  $u_\tau$       &    5.13            \\
        5b      &  $u_\tau^*$     &    12.4            \\
        5c      &  $u_\tau^*+u_q$ &    4.18            \\ \hlineB{3}
    \end{tabular}
    \caption{Summary of Method 5}
    \label{tab:mtd-5-summary}
\end{table}

\begin{figure}[htbp]
    \begin{center}
        \begin{subfigure}{\textwidth}
            \begin{center}
                \includegraphics[width=0.485\textwidth,trim={3.4cm 8.5cm 3.4cm 8.5cm},clip]{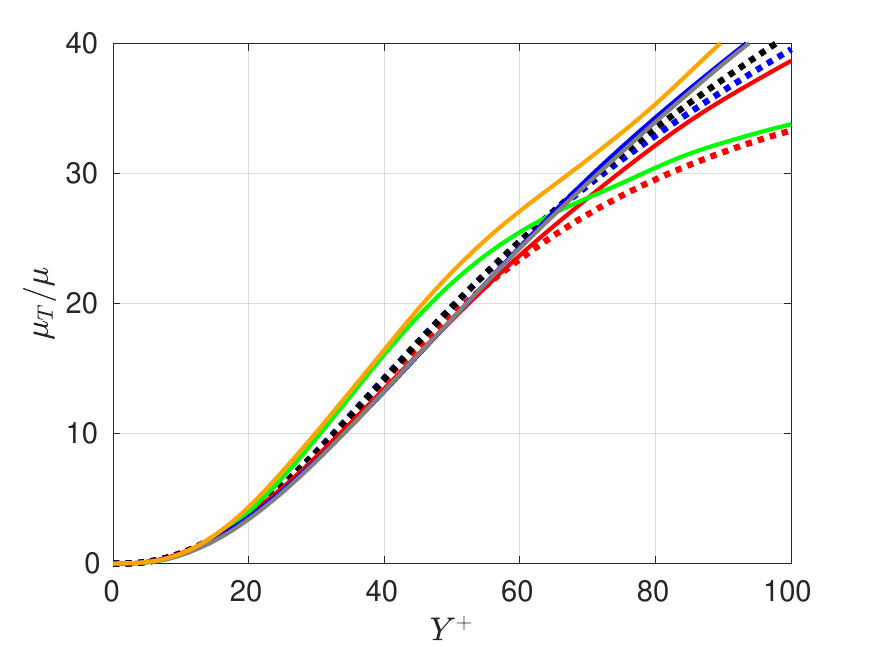}
                \includegraphics[width=0.485\textwidth,trim={3.4cm 8.5cm 3.4cm 8.5cm},clip]{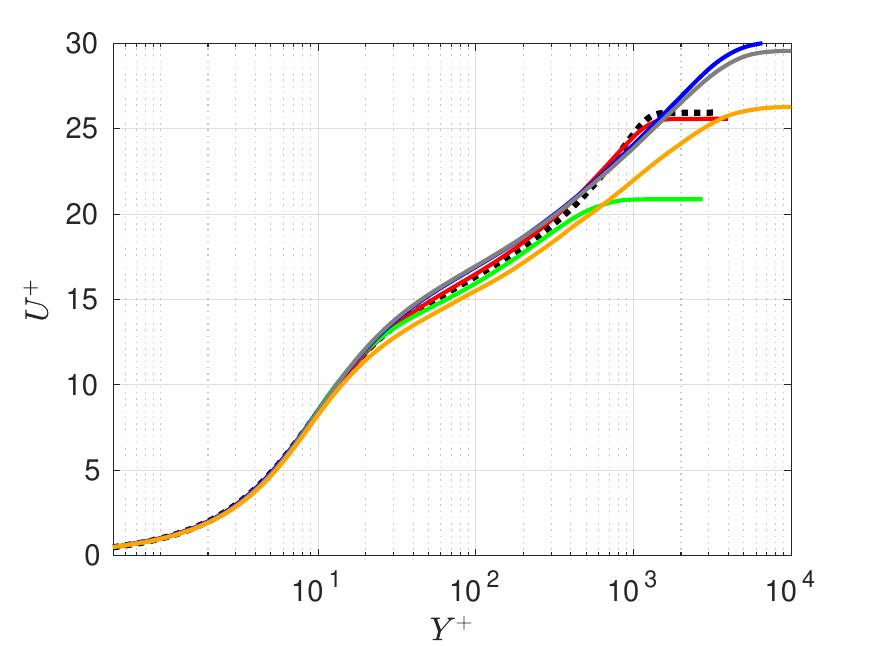}
                \caption{Method 5a $u_c=u_\tau$}
            \end{center}
        \end{subfigure}
        \begin{subfigure}{\textwidth}
            \begin{center}
                \includegraphics[width=0.485\textwidth,trim={3.4cm 8.5cm 3.4cm 8.5cm},clip]{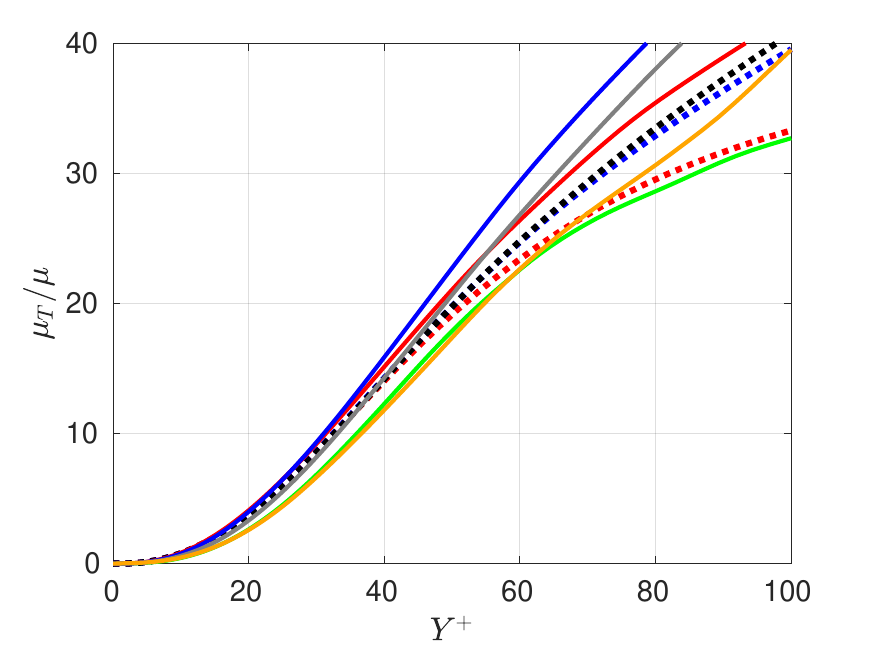}
                \includegraphics[width=0.485\textwidth,trim={3.4cm 8.5cm 3.4cm 8.5cm},clip]{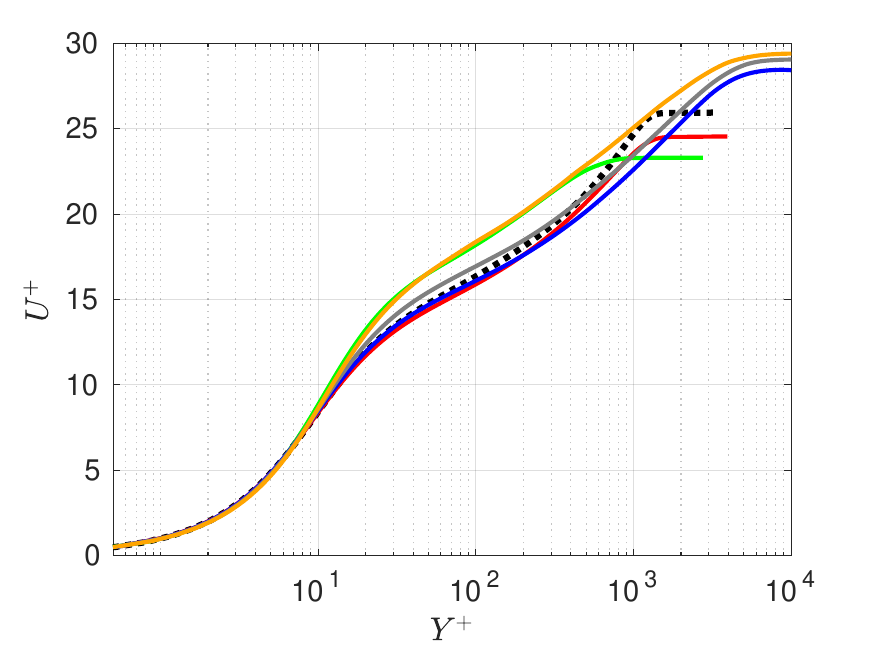}
                \caption{Method 5b $u_c=u_\tau^*$}
            \end{center}
        \end{subfigure}
        \begin{subfigure}{\textwidth}
            \begin{center}
                \includegraphics[width=0.485\textwidth,trim={3.4cm 8.5cm 3.4cm 8.5cm},clip]{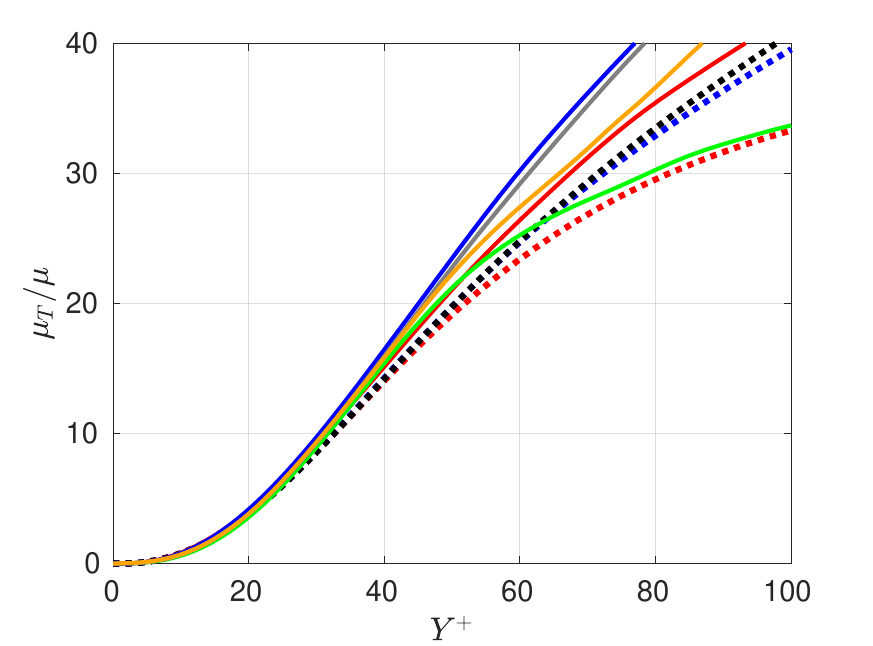}
                \includegraphics[width=0.485\textwidth,trim={3.4cm 8.5cm 3.4cm 8.5cm},clip]{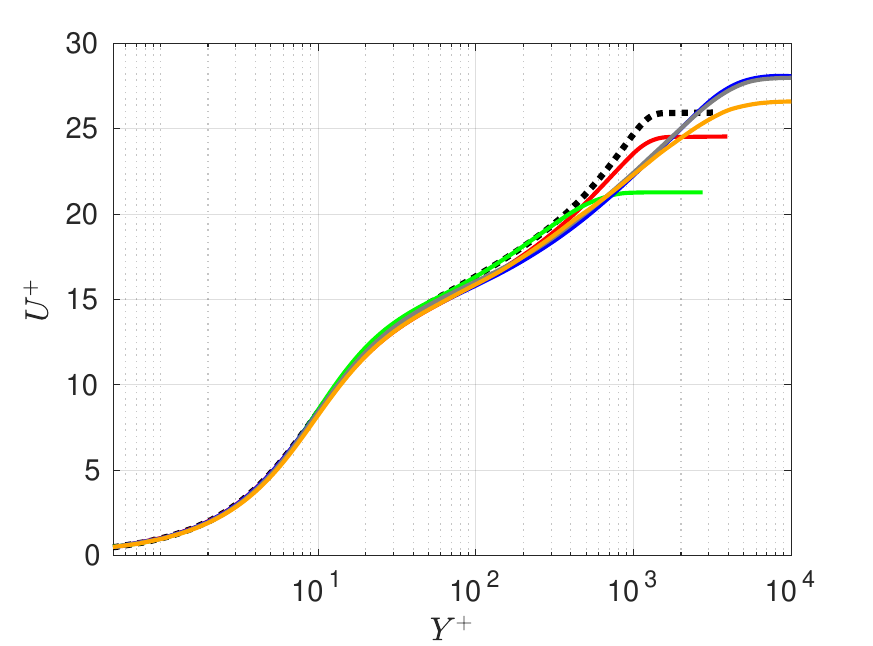}
                \caption{Method 5c $u_c=u_\tau+u_q$}
            \end{center}
        \end{subfigure}
    \end{center}
    \caption{Transformed Profiles with Method 5. Eddy viscosity is on the left and velocity is on the right. Legends are the same as those in \Cref{fig:profiles-other-methods}.}
    \label{fig:mtd-5}
\end{figure}

\Cref{fig:mtd-5} shows the influence of $u_c$ choices on the transformations. Eddy viscosity profiles obtained from Method 5a with $u_c=u_\tau$ have slightly higher sensitivity to wall cooling than the transformation in Volpiani et al. \cite{volpiani2020}, which results in a larger velocity error of $5.13\,\%$ in the inner layer. Interestingly, Method 5b with $u_c=u_\tau^*$ has a larger variation in eddy viscosity and this causes velocity errors up to $12.4\,\%$. By adding  $u_q$, to include wall heat transfer, Method 5c improves the collapse of  eddy viscosity profiles for $Y^+\lessapprox40$, which mostly removes the velocity errors of Method 5b. 

\subsection{Method 6 (Iterative transformation based on inner/outer scaling)}
Inspired by \cite{hasan2024}, the last method transforms the whole boundary layer by using both inner and outer scaling of incompressible flows. For that reason, it deviates from the other approaches. Unlike \cite{hasan2024}, however, we propose a transformation via an iterative scheme, without employing additional compressibility treatments for the inner and outer scaling of incompressible flows.    

In each iteration, the proposed transformation is determined by
\begin{equation}
    Y(y) = 
    \begin{cases}
        Y_1(y), & \text{for } Y<0.2\delta_{99,inc}\\
        Y_2(y), & \text{for } 0.2\delta_{99,inc}\le Y\le\delta_{99,inc}
    \end{cases}
\end{equation}
where $Y_1$ is obtained from the incompressible mixing length model and Van Driest damping, as in Method 2, while the law of the wake with Cole's wake function,
\begin{equation}
    U(Y) = \frac{u_\tau}{\kappa}\log{\left(\frac{Yu_\tau}{\nu_w}\right)}+Bu_\tau+2\frac{\Pi u_\tau}{\kappa}\sin^2{\left(\frac{\pi Y}{2\delta_{99,inc}}\right)},
\end{equation}
is used estimate $Y_2$.
For present purposes, the wake parameter is the zero-pressure gradient value $\Pi=0.8\times(0.5^{3/4})$. The wall-normal derivative of incompressible velocity is given as
\begin{equation}
    \frac{dU}{dY}=\frac{u_\tau}{\kappa Y}+\frac{\Pi u_\tau}{\kappa\delta_{99,inc}}\sin{\left(\frac{\pi Y}{\delta_{99,inc}}\right)}
\end{equation}
By the velocity scaling (\ref{eq:vel-grad-hypothesis}), we obtain a nonlinear equation for $Y_2$
\begin{equation}
    \begin{aligned}
        \frac{u_\tau}{\kappa Y_2}+\Pi \frac{\pi u_\tau}{\kappa\delta_{99,inc}}\sin{\left(\frac{\pi Y_2}{\delta_{99,inc}}\right)}-\frac{\mu}{\mu_w}\frac{du}{dy}=0
    \end{aligned}
\end{equation}
Since $(\mu/\mu_w)(du/dy)$ is available from a compressible DNS, this relation can be solved by Newton's method for $Y_2$.

Note that this transformation does not use incompressible DNS data in its formulation. In addition, unlike the previous approaches where the upper bound of the inner layer is taken from compressible coordinates as $0.2\delta_{99}$, this method sets it from the transformed incompressible coordinate as $0.2\delta_{99,inc}=0.2Y(\delta_{99})$. The iterations are assumed to be converged when $\delta_{99,inc}$ is sufficiently converged.

\begin{figure}[htbp]
    \begin{center}
        \begin{subfigure}{0.485\textwidth}
            \centering
            \includegraphics[width=\textwidth,trim={3.4cm 8.5cm 3.4cm 8.5cm},clip]{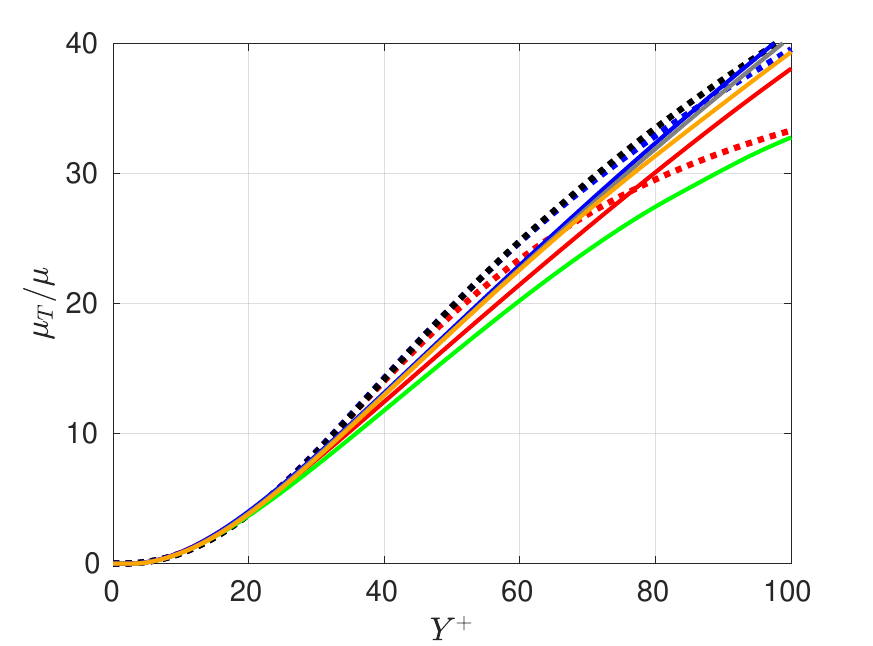}
            \caption{Eddy Viscosity}
        \end{subfigure}
        \begin{subfigure}{0.485\textwidth}
            \centering
            \includegraphics[width=\textwidth,trim={3.4cm 8.5cm 3.4cm 8.5cm},clip]{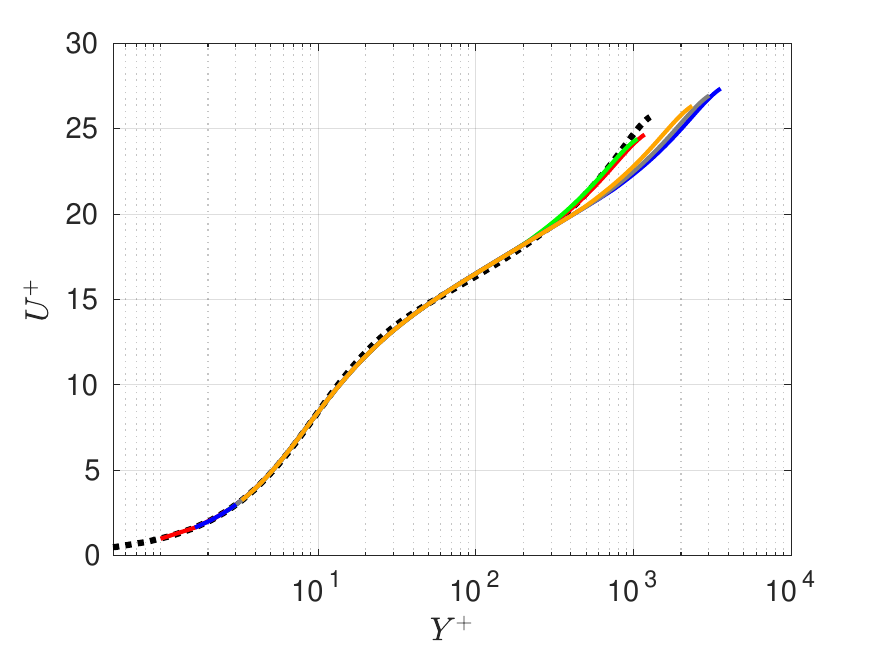}
            \caption{Velocity}
        \end{subfigure}
        \begin{subfigure}{0.485\textwidth}
            \centering
            \includegraphics[width=\textwidth,trim={3.3cm 8.5cm 3.3cm 8.5cm},clip]{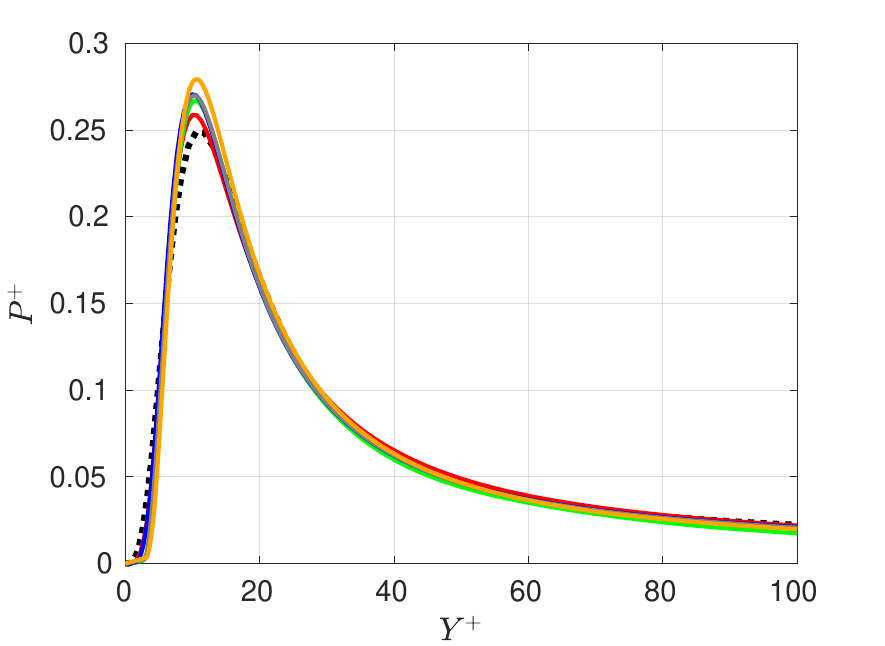}
            \caption{TKE Production}\label{fig:mtd-6c}
        \end{subfigure}
        \begin{subfigure}{0.485\textwidth}
            \centering
            \includegraphics[width=\textwidth,trim={3.3cm 8.5cm 3.3cm 8.5cm},clip]{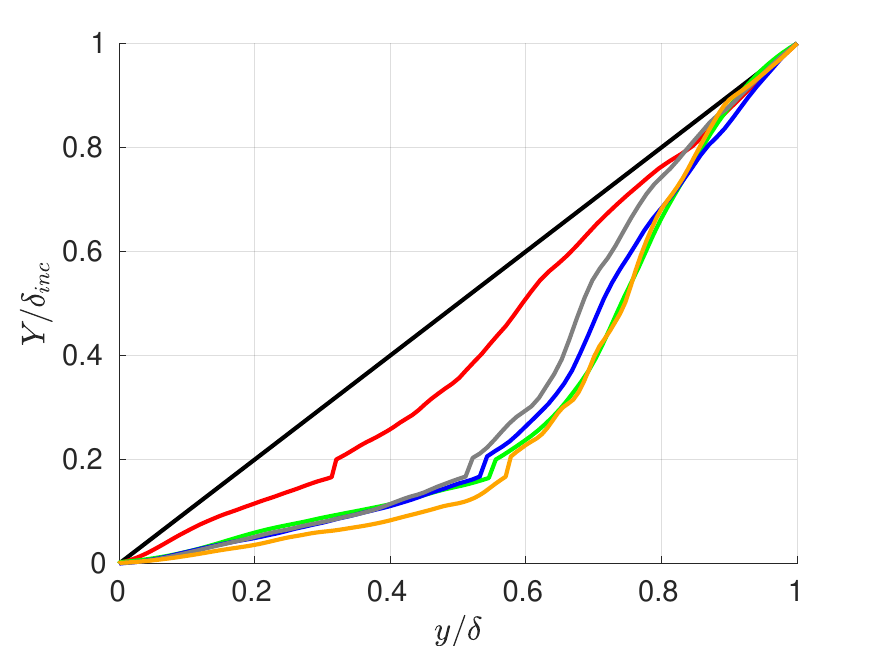}
            \caption{Transformed wall-normal coordinate}\label{fig:mtd-6d}
        \end{subfigure}
    \end{center}
    \caption{Transformed Profiles with Method 6. Legends are the same as those in \Cref{fig:mtd-1}.}
    \label{fig:mtd-6}
\end{figure}

\Cref{fig:mtd-6} shows the profiles of transformed eddy viscosity, velocity and TKE production. Compared to Method 2, Method 6 removes the kinks and results in smooth eddy viscosity profiles due to better matching conditions at $Y=0.2\delta_{99,inc}$. The velocity profiles are shown only up to the boundary layer edge, as this transformation holds only inside the boundary layer. We observe that the compressible velocity profiles collapse on incompressible profiles accurately, with a maximum error of $1.93\%$. Note that this error is measured for $y<0.2\delta_{99}$, which is consistent with the previous cases and also with the fact that the logarithmic layer of the actual incompressible data is shorter compared to that of the transformed versions of M6Tw076, M8Tw048 and M14Tw018 cases. \Cref{fig:mtd-6d} demonstrates the most notable difference of the present method to previous ones. While high Mach numbers push the transform wall-normal coordinate profiles below the incompressible line, all hypersonic cases have similar trends while the profiles of the cold wall cases (M6Tw025 and M14Tw018) almost collapse on the other. This might be because the lower bound of the outer layer is taken from the incompressible state and it is determined iteratively. Larger variations in the material properties due to larger variations in the temperature in these cases might also have been another factor.

\section{Conclusion}
In this paper, we showed the significance of satisfying the eddy viscosity equivalence below the logarithmic layer for deriving accurate compressibility transformations. We demonstrated that the Trettel and Larsson \cite{trettel2016} transformation does not meet this criterion, while the Volpiani transformation \cite{volpiani2020} is better, except for a slight sensitivity to wall cooling, which makes it a more accurate transformation. To remove the dependence on wall-cooling, we introduced two approaches -- curve fitting incompressible DNS eddy viscosity profiles and transforming incompressible Van Driest damping. We proposed several new transformations that use these ideas, including one that significantly improves the accuracy of the Volpiani transformation \cite{volpiani2020}. This success motivated a new integral transformation, which introduced a new compressible velocity scale. We showed that accurate transformations can be obtained by respecting the physics of the turbulence as well as compressibility and wall cooling effects.

\section*{Acknowledgments}
This work was funded in part by the Office of Naval Research grant \#N00014-21-1-2264.

\clearpage
\bibliographystyle{plain}   
\bibliography{references}

\end{document}